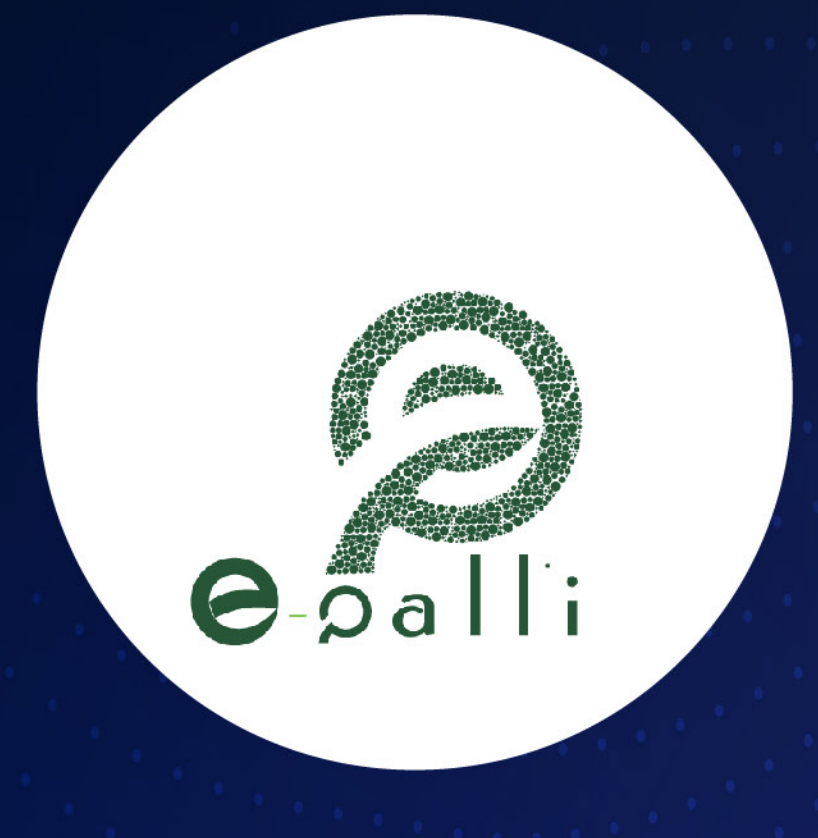

# Applied Research and Innovation (ARI)

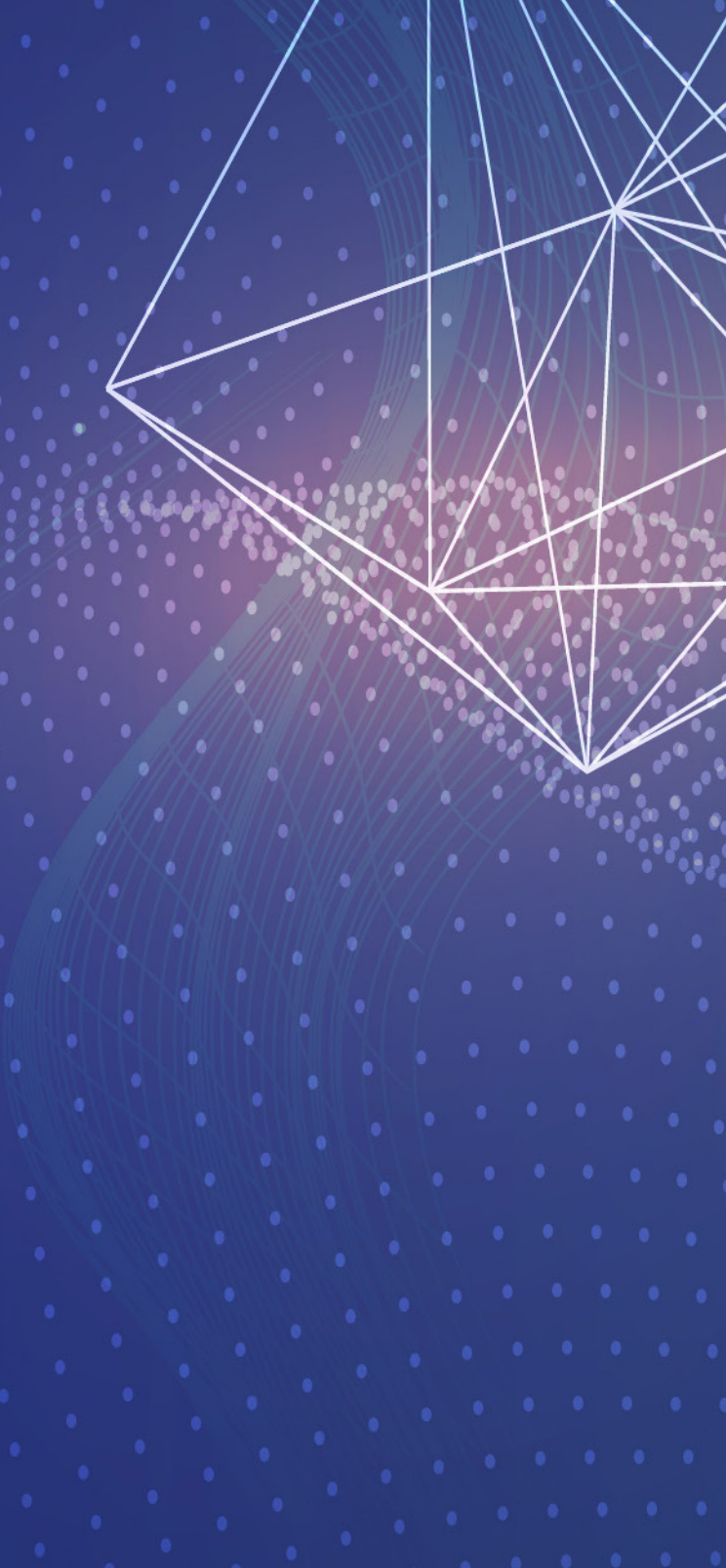

# So Many Opinions, So Many LLMs: Comparing Large Language Models to Traditional Machine Learning for Open- Ended Survey Analysis

Abdullah Akinde[1*], Mariam Akinde[1], Rasheedat Emiola[1], Ahmed Akinsola[1]

## ABSTRACT

Open-ended surveys offer valuable insights, but they are notoriously difficult to analyze at scale. Building on previous work that employed traditional machine learning to classify text ("So Many Responses, So Little Time: A Machine-Learning Approach to Analyzing Open-Ended Survey Data"), this study investigates how different large language models (LLMs) understand and analyze NSSE open-ended survey responses. We focus on several cutting-edge LLMSs, including OpenAI's GPT-4-Turbo, Claude 3.5, Meta's LLaMA 3 (70B), and Perplexity Sonar Pro, and compare their performance to the previous machine learning models in tasks like sentiment analysis and thematic classification.
Our research analysis assesses model agreement, classification accuracy, and interpretability of reasoning. The findings reveal that current LLMs routinely beat classic machine learning models in classification accuracy, particularly in understanding complex mood and theme patterns in student replies. While LLMs have superior accuracy, they differ greatly in how explicitly and consistently they justify their predictions and apply category boundaries. These distinctions highlight crucial trade-offs when using LLMs for qualitative analysis: greater predictive power comes with challenges of consistency and explainability. Our findings illustrate the benefits and drawbacks of utilizing various LLMs for large-scale qualitative research, and we provide practical advice for researchers looking to balance automation and interpretive rigor.

## INTRODUCTION

Open-ended replies in the NSSE survey provide a wealth of qualitative information that supplements quantitative findings. Making sense of such large amounts of material, however, is a chronic challenge. In our previous research, "So Many Responses, So Little Time: A Machine-Learning Approach to Analysing Open-Ended Survey Data" (A. Michael & A. Abdullah 2024), we utilized traditional natural language processing (NLP) and machine learning techniques to partially automate thematic coding and sentiment analysis. However, these models are often based on substantial feature engineering and may fail to reflex the complexities and ambiguities of the human language.

This limitation is most apparent in emotion classification, when surface level polarity (positive, negative, or neutral) is not sufficient. The typical Survey responses frequently include subtle emotional states, such as dissatisfaction masked as sarcasm, thankfulness laced with exhaustion, or hope tempered by skepticism, that require pragmatic inference and world knowledge to understand. Traditional pipelines cannot handle such contextual disambiguation, resulting in misclassification of complex expressions (Pang, B., & Lee, L. 2008; Liu, 2015). Moreover, domain specific conventions such as academic discourse markers in NSSE data ("The advising was fine, though…" implying discontent), might complicate rule based systems. To address this gap, recent computational work has turned to psychologically grounded emotion frameworks. Instead of contrasting positive and negative feelings, researchers have begun viewing sentimental analysis as forms of recognizing a range of emotions that can be interpreted by people (Bostan & Klinger, 2018; Saravia, *et al.*, 2018). This view aligns better with how humans feel and allows for more detailed analysis which can guide action by institutions when responses come from emotions.Thus, any automated system must be evaluated not only for accuracy, but also for consistency and bias robustness across diverse respondent subgroups. Recent advances in large language models (LLMs) have provided us with opportunities for evaluating open-ended data. Unlike previous models, contemporary LLMs can do context-sensitive reasoning, abstraction, and emotion recognition. We are using them to determine which of the 6 basic emotions is the response. However, since LLMs differ in their architectures, training data, and alignment aims, their outputs may differ in small but significant ways. This raises an important question: do these models genuinely agree about how humans feel?

This study addresses that question by comparing the performance of various top LLMs on a set of open-ended survey replies. Our goal is to determine how consistent their sentiment readings are and where they differ. In doing so, we hope to contribute to larger arguments concerning LLM dependability, interpretability, and their role in furthering computational social science and user experience research.

[1] Austin Peay State University, United States
[*] Corresponding author's e-mail: aakinde@my.apsu.edu

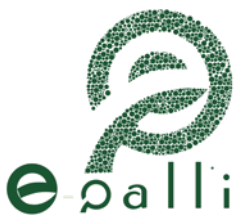

## LITERATURE REVIEW

Natural Language Processing (NLP), which refers to a machine's ability to convert human speech into a computable format, has undergone a dramatic change over the last seventy years, going through three primary phases: first was Symbolic NLP (where computers mimic human thought in a defined set of rules), then Statistical NLP (where computers statistically predict the likelihood of a certain response based on previous responses), and finally Neural NLP (where computers utilize deep learning models to train themselves to generate "natural" language responses) (Jurafsky & Martin, 2026). Each phase in the NLP developmental path has occurred as machine learning techniques have progressed, as the amount of available data to train trained machine learning models has increased, and as computers have become more capable of processing large amounts of information rapidly.

The fact that NLP has transitioned from being based on fixed rules to being based on dynamic adaptive self-learners shows that it has developed into an integral part of Artificial Intelligence that can not only analyze language but also perform both practical reasoning and action based upon language analyses.

The development path of NLP is why Sentiment Analysis has emerged as a benchmark for testing language understanding and why LLMs are currently pushing the limits of their affective and contextual understanding of language.

### Sentiment Analysis

Sentiment Analysis (SA), a core subfield of Natural Language Processing (NLP), seeks to computationally identify extract and quantify subjective context within textual data, including opinions, attitudes, evaluations, and emotions (Liu, 2015; Poria *et al.*, 2020)., frequently derived from feedback on websites, social media, and in our case open-ended survey responses. The goal is to develop and train computational models that can infer sentiment with human-like accuracy, with the potential to be scaled up for a vast dataset using cloud-based inference.

This approach builds on previous work that used supervised learning on unstructured text data, like classifying open-ended survey responses to handle high-volume qualitative input. It then moves on to more detailed, context-aware modelling of emotions and feelings. Furthermore, the modelling of the entire range of human subjective experience encompassing emotions, attitudes, and value-laden judgments through sophisticated sentiment analysis constitutes a pivotal achievement in the pursuit of Artificial General Intelligence (AGI), as it facilitates machines in understanding and reacting to the affective and evaluative aspects of human communication (Bubeck *et al.*, 2023).

### Large Language Models

In recent years, large language models transitioned away from being basic statistical learners into tool- enhanced reasoning agents, thus, they have undergone a dramatic transformation. Foundation models like GPT-3 (Brown *et al.*, 2020), PaLM (Chowdhery *et al.*, 2022), and LLaMA (Touvron *et al.*, 2023) have successfully trained on massive volumes of text, and also see improved performance through Instruction tuning (Wei *et al.*2020) and Reinforcement Learning of Human-Feedback (RLHF) (Christiano *et al.*, 2017). Consequently, their ability to perform well in both zero-shot and few-shot settings has prompted the transition from the traditional practice of employing task-specific finetuning on models to the novel application of prompt-based approaches for model adaptation.

The above changes in LLMs are reflected in veterinary literature that uses sentiment analysis (SA) for the following examples: Zhong *et al.* (2023) compared the performance of zero-shot models against the performance of fine-tuned BERT on standard benchmark tasks, Wang *et al.* (2023) assessed the ability of ChatGPT to process shifting polarity and make open-domain inferences and finally Deng *et al.* (2023) used LLMs to help create weak supervision and achieved almost the same performance of fully supervised baselines at the same time. However, these studies largely predate the current generation of reasoning- optimized models.

In 2024–2025, architectures such as GPT-4o, Claude 3.5 Sonnet (Anthropic, 2024), o1 (OpenAI, 2024), and LLaMA-3 (Meta AI, 2024) have redefined the frontier, combining multimodal grounding, ultra-long-context reasoning (>10M tokens), self-verification, and native tool use, often trained with scalable alternatives to RLHF (e.g., Direct Preference Optimization [DPO]; Rafailov *et al.*, 2023). These capabilities enable structured, auditable reasoning, essential for complex SA tasks involving domain-specific constraints (e.g., financial tone classification under SEC/FINRA compliance protocols) or real-time adaptation to market sentiment dynamics. As a result of this change from static prompting to Interactive Agentic Reasoning, LLMs (Large Language Models) are now being utilized as dynamic virtual Agents/Components within Autonomous Systems That Are Capable of Developing and Executing Plans, Self-Correcting, Making Compliance Aware Decisions with a Built in Understanding of Regulatory Requirements.

Even with the rapid progression of technology, the availability of LLMs for evaluating real-world SA (Situational Awareness) Tasks where temporal coherence, explanation, and regulatory alignment are critical factors still remains limited.

### Traditional Machine Learning Models

Traditional machine learning (ML) methods, which continue to fulfill a critical role even now with the introduction of recent advancements in deep learning and large language models (LLMs), are particularly well suited for use when the characteristics of the data are clearly defined, interpretability must be enforced, or when speed in computing is of utmost importance (Jordan & Mitchell, 2015; Domingos, 2012). The majority of the ML methods referenced in previous research are based on supervised learning techniques and need to rely on the creation of explicit model features, through such processes as Extracting Term Weighting by Inverted Document Frequency (TF-IDF), tokenising text into n-grams and creating handcrafted lexical– syntactic features. These methods can provide an excellent baseline for performance, especially for classification tasks, such as determining sentiment (Pang & Lee, 2008), and therefore, can be a useful starting point for creating a high-quality classifier. In terms of classifying sentiment as an example, the Support Vector Classifier (SVC) with an RBF kernel, also known as "kernelized SVC," creates sophisticated, non-linear decision boundaries through implicit mapping into a high dimensional space, maximising the margin between classes while simultaneously controlling overfitting using hinge loss and $\ell_2$ regularisation; SVC with RBF kernel has been shown to be effective even when using representation methods that produce sparse, high-dimensional text representations (Cortes & Vapnik, 1995; Joachims, 1998). Because of this scenario, the LinearSVC, the linear version of the Support Vector Classifier, is less expressive, though it is faster to train due to the use of primal optimisation (e.g., LIBLINEAR) and therefore can provide a solution for users with a need to rapidly develop sentiment classifiers using high-volume text datasets (Fan *et al.*, 2008; Wang & Manning, 2012).

In addition, Random Forest, a classification machine-learning algorithm, is able to overcome regularization, variance, and generalization through an ensemble approach by using both random sampling and aggregated sample groups of unknown sizes in order to increase accuracy while providing interpretability through the calculation of feature importance. This is very important in regulated industries such as finance or healthcare. Finally, Decision Trees are a type of classification algorithm that provide human-readable, easily understandable decision-making processes and can be further regularized by controlling their depth or through techniques of pruning and help ensure compliance in highly regulated environments, where it is critical to maintain compliance by maintaining easy auditability through clear integration into legally established rules-based systems. Together, all three classification algorithms provide an outstanding balance of accuracy, efficiency, and explainability.

## MATERIALS AND METHODS

This research paper presents an approach to assess the comparative performance of traditional Machine Learning (ML) methods vs. state-of-the-art Large Language Models (LLMs) for a multi-class emotion recognition task. To evaluate performance, we will use three separate text datasets - two cohorts from the National Survey of Student Engagement (NSSE; 2019 and 2021) and the Hugging Face Emotion Dataset. These diverse dataset sources should give us a unique perspective on how models perform across different domains, sample sizes, label distributions and therefore help to evaluate robustness of the models. The proposed approach has 4 major steps: (1) Dataset Curation & Pre-Processing (2) Classical ML Baseline Implementation (3) Foundation LLMs Zero-shot Prompt Inference and (4) Standardised Evaluation Methods based on Accuracy and per class Prediction Counts. Ultimately, we anticipate that this approach will enable us to provide an in-depth evaluation of LLM capabilities for Fine-Grained Sentiment Analysis while minimising the effects of extreme class imbalance and data leakage between training and evaluation datasets. Text and graphic files should be kept separate until after formatting and styling.

### NSSE Datasets

The National Survey for Student Engagement (NSSE) is an annual survey that gathers information about how U.S. colleges and universities engage their undergraduate students through their educational practices. The current study analysed two derived datasets which were available from the NSSE's datasets: 2019 and 2021 for a university of this size. As part of the analysis each dataset was processed to remove any blank responses that were not collected by the surveys. After processing both datasets, a final number of annotated sample sizes was generated: 2021 = 293 and 2019 = 142.

### Hugging Face Emotion Data

The Hugging Face Emotions dataset contains English language tweets that were selected based on their 6 main categories of emotion (anger, fear, joy, love, sad, and surprise). These datasets were divided into training sets of 70,0004 tweets and a testing set of 17,950 tweets using a stratified sampling method in order to maintain the overall population distribution.

### Classical ML Baseline

All classical ML models were trained exclusively on the Hugging Face Emotion training set (N=70,004) and evaluated in a zero-shot fashion on the NSSE responses—i.e., no NSSE data was used during training or validation. We implemented four classical machine learning (ML) classifiers as baselines:

### Support Vector Classifier (SVC):

The Support Vector Classifier using Radial Basis Function kernel and trained on a sample set of 20,000 achieved an accuracy of 90.08%, which is nearly equal to LinearSVC using only a small subset of data compared to LinearSVC. Thus, non-linear decision boundaries may provide incremental benefits, as it appears to be able to learn nuanced lexical and contextual relationships ignored by linear models (for instance, negation and sarcasm). However, the computation expense associated with the RBF-SVC increases exponentially with respect to the size of the data set, which results in reduced feasibility when using large data sets or in the context of real-time inferences without using approximations.

```
SVC (Subset) Test Accuracy: 0.8552
Classification Report:

              precision    recall  f1-score   support

     sadness       0.83      0.81      0.82       293
         joy       0.79      0.79      0.79       295
        love       0.85      0.93      0.88       299
       anger       0.85      0.87      0.86       303
        fear       0.92      0.77      0.84       296
    surprise       0.90      0.96      0.93       309

    accuracy                           0.86      1795
   macro avg       0.86      0.85      0.85      1795
weighted avg       0.86      0.86      0.85      1795
```

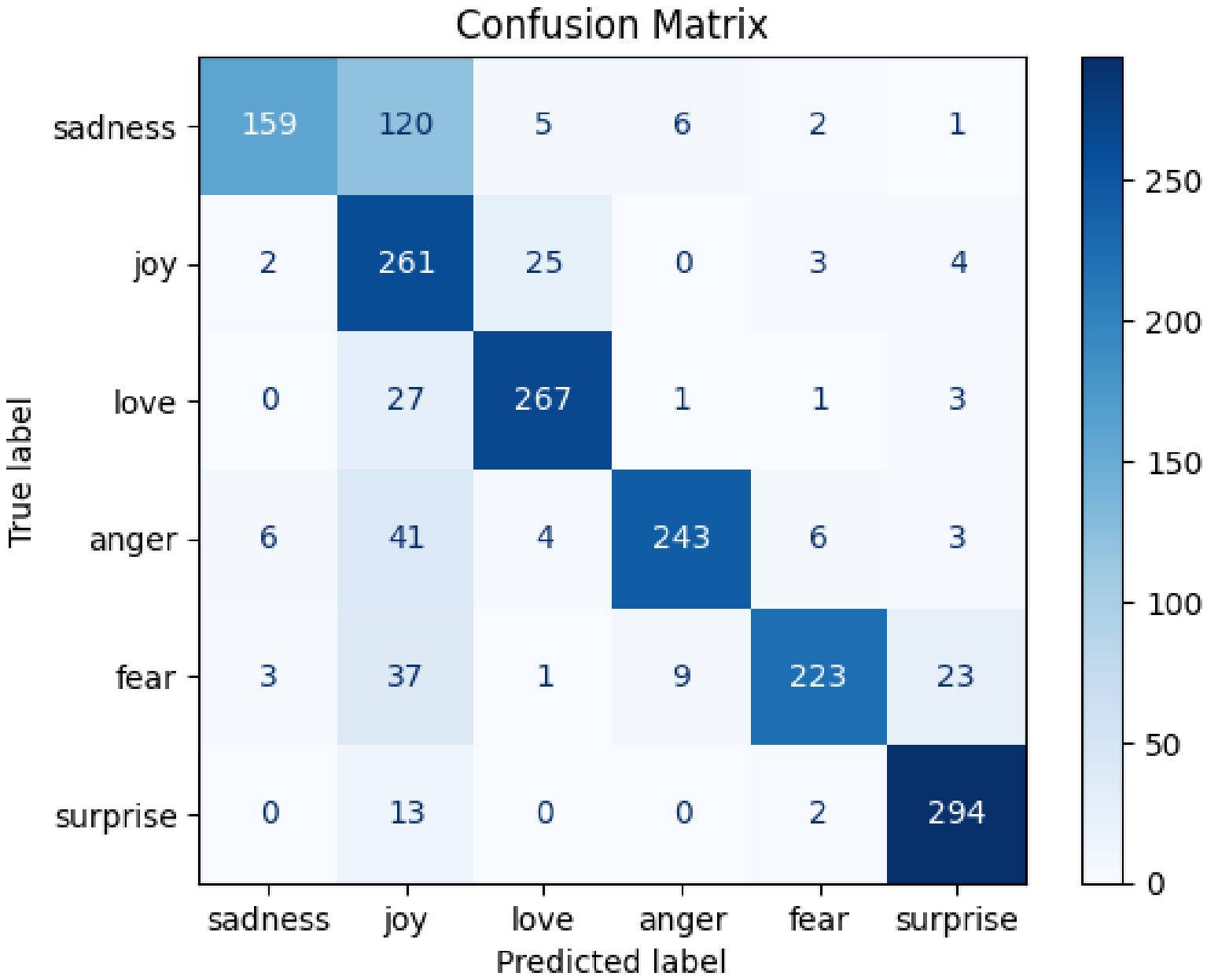

**Figure 1**: SVC Subset Accuracy and Confusion Matrix

**Linear Support Vector Classifier (LinearSVC):**
The LinearSVC achieves the best accuracy score of 90.98% with strong performance across all classes indicated by the F1-scores which range from 0.87 to 0.93. The combination of a linear model and efficient optimization techniques used in conjunction with the ability to train on fully data sets makes the LinearSVC both highly accurate and highly scalable. Therefore, the LinearSVC is an excellent choice for production applications where speed, ease of auditing (by inspecting weights), and legal compliance are critical goals.

The Random Forest with 50% of the available data (5,000 features using richer TF-IDF and bigrams) and only 20 trees is encouragingly very close to achieving 89.78% accuracy which is almost identical to RBF- SVC and nearly equal to LinearSVC. The Random Forest also has a macro F1 of 0.90 meaning it has equal performance across all six emotions. In particular, the Random Forest has the feature of built-in feature importance and therefore continues to provide a level of explainability, which is valuable when explainability and robustness are almost as critical as accuracy (for example, within regulated autonomous systems, where all decisions need to be justified).

**Decision Tree Classifier:**
Decision Tree Classifier on the entire data set, the performance was marked by 85.48% accuracy with substantial decreases in precision for the classes of fear (79%) and joy (82%). Although it provides a degree of interpretability, allowing one to precisely reconstruct the series of decisions made, its simplicity leads to a poor ability to model the high-dimensional nature of text features. As such, it serves as a useful baseline and provides diagnostic capabilities; however, it cannot be considered a leading solution.

In summary, the efficiency winner continues to be LinearSVC, while RBF-SVC demonstrates an advantage for non-linearity (albeit at a cost), and a scaled Random Forest provides a balance between performance and interpretability, making it an excellent option for transparent and high-stakes applications.

```
LinearSVC Test Accuracy: 0.9098
Classification Report:

              precision    recall  f1-score   support

     sadness       0.93      0.91      0.92      2991
         joy       0.93      0.87      0.90      2992
        love       0.90      0.95      0.93      2992
       anger       0.92      0.92      0.92      2992
        fear       0.89      0.85      0.87      2992
    surprise       0.89      0.95      0.92      2992

    accuracy                           0.91     17951
   macro avg       0.91      0.91      0.91     17951
weighted avg       0.91      0.91      0.91     17951
```

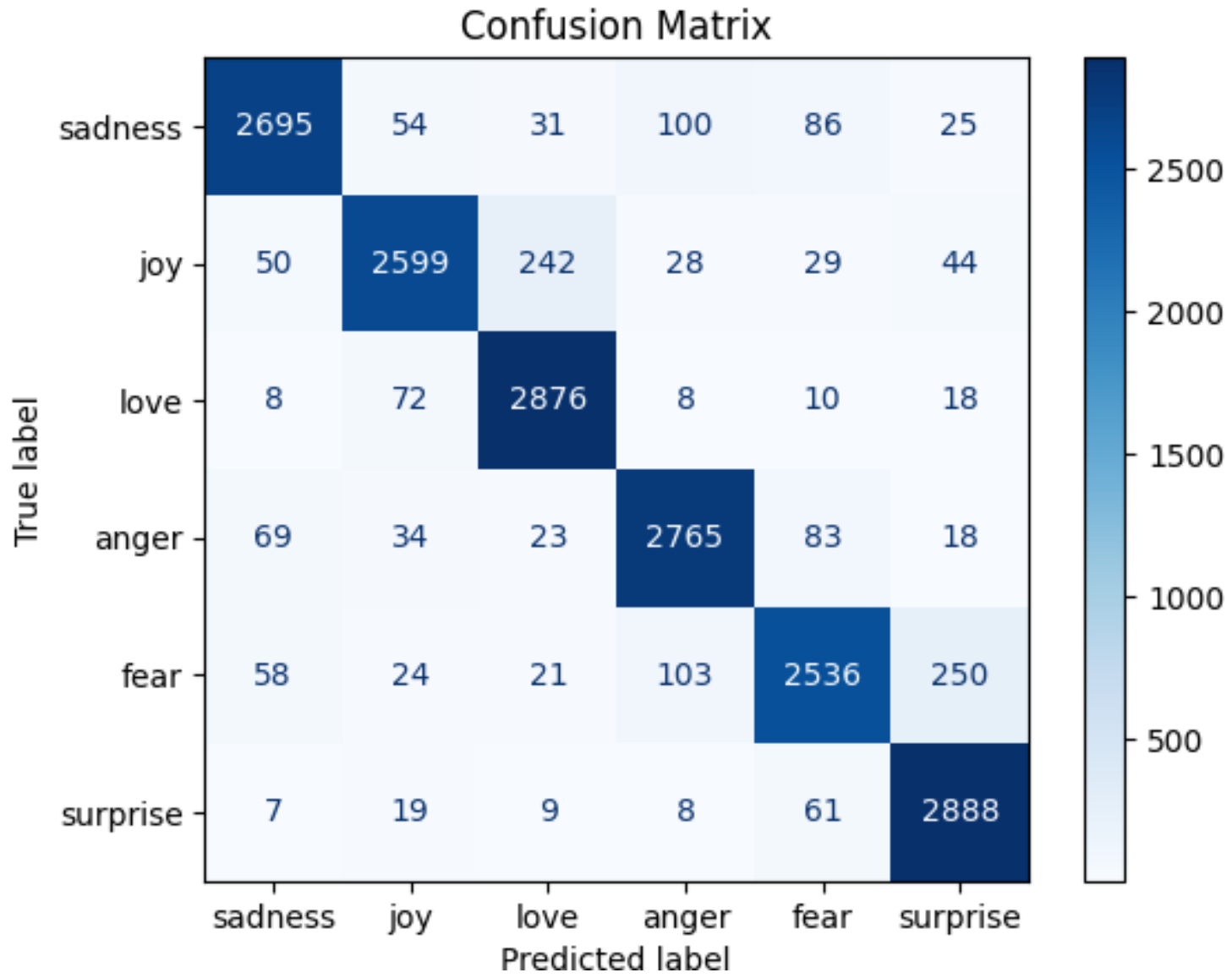

**Figure 2**: Linear SVC Subset Accuracy and Confusion Matrix RandomForestClassifier.

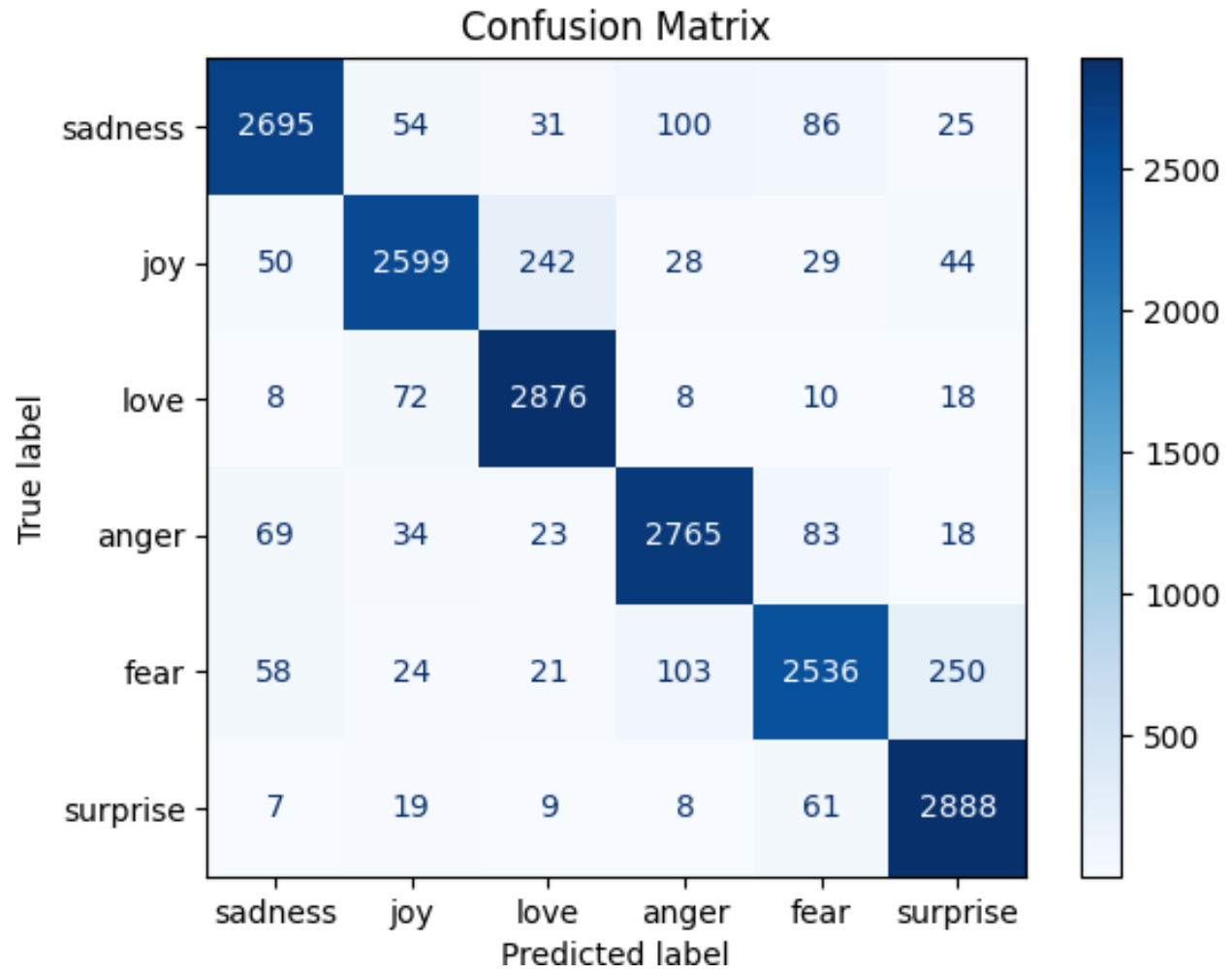

**Figure 3**: Random Forest Classifier Subset Accuracy and Confusion Matrix

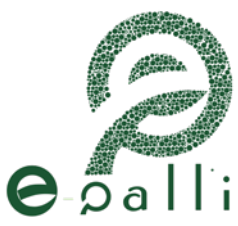

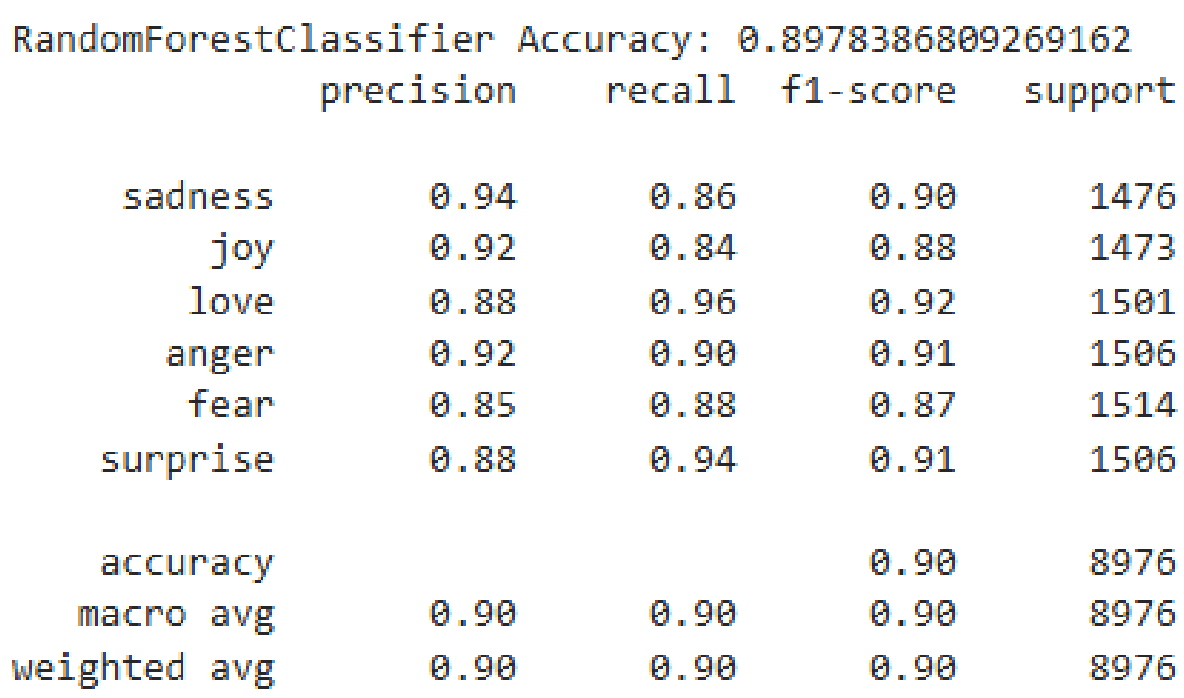

RandomForestClassifier Accuracy: 0.8978386809269162

| | precision | recall | f1-score | support |
|---|---|---|---|---|
| sadness | 0.94 | 0.86 | 0.90 | 1476 |
| joy | 0.92 | 0.84 | 0.88 | 1473 |
| love | 0.88 | 0.96 | 0.92 | 1501 |
| anger | 0.92 | 0.90 | 0.91 | 1506 |
| fear | 0.85 | 0.88 | 0.87 | 1514 |
| surprise | 0.88 | 0.94 | 0.91 | 1506 |
| accuracy | | | 0.90 | 8976 |
| macro avg | 0.90 | 0.90 | 0.90 | 8976 |
| weighted avg | 0.90 | 0.90 | 0.90 | 8976 |

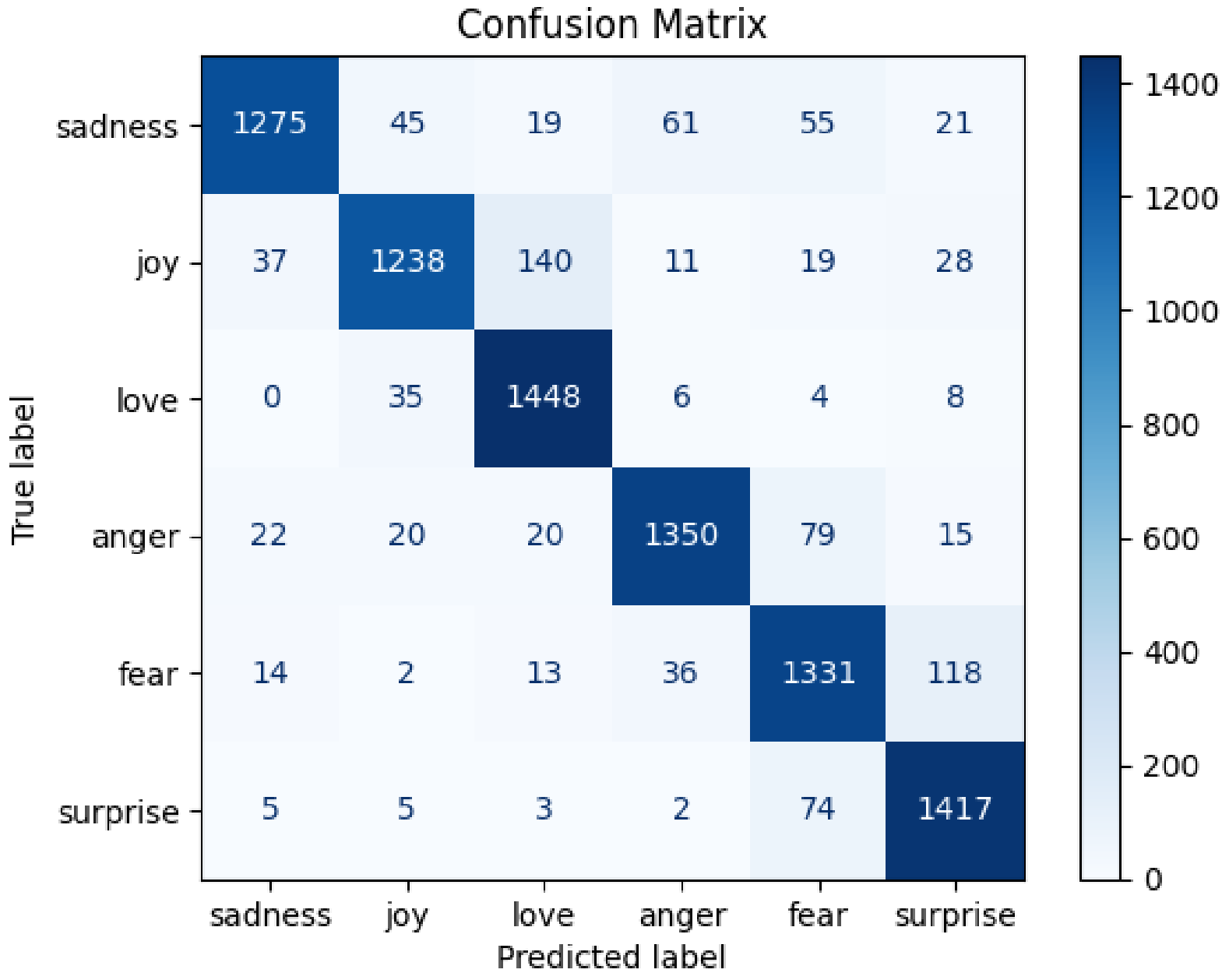

**Figure 4**: RandomForestClassifier Subset Accuracy and Confusion Matrix

**LLM Prompting & Evaluation**

Language Models (LLMs) were evaluated in zero-shot mode without fine-tuning on our datasets. We selected four state-of-the-art LLMs:

- OpenAI GPT-4-Turbo (gpt-4-0125-preview)
- Claude 3.5 Sonnet (Anthropic)
- Meta LLaMA 3 (70B) (open-source, locally deployed via llama.cpp)
- Perplexity Sonar Pro (Perplexity AI's instruction-tuned model)

Each LLM was prompted with the following template:
"Take the text given below, and classify the emotion in the text into one of the following six categories: Joy, Surprise, Love, Anger, Sadness, Fear. The following is the input text '{input_text}' → Emotion:"
Responses were parsed using exact string matching (case-insensitive) to assign predicted labels. To ensure reproducibility, all API calls used temperature = 0 and max_tokens = 5. For local models (LLaMA 3), we used greedy decoding. Predictions were aggregated per emotion class to compute confusion matrices and per-class counts. Since LLMs are not trained on our specific datasets, their Training Accuracy is undefined and reported as "–" in all tables.

## RESULTS AND DISCUSSION

This section presents a comparative evaluation of conventional machine learning (ML) models and contemporary large language models (LLMs) on both benchmark sentiment classification and real-world open-ended survey data. All experiments employed identical preprocessing (e.g., lowercasing, punctuation normalization, and removal of NSSE identifiers) and used stratified train–test splits where applicable. Crucially, no fine-tuning was performed on the LLMs; evaluations were conducted via zero-shot prompting with standardized emotion classification instructions grounded in the six basic emotions framework.

**Benchmark Performance: Accuracy and Generalization Gap**

The traditional ML models evaluated show good generalization performance on this balanced classification of emotions task, with very limited train–test separation (which is counter-intuitive to the overfitting issue usually associated with text modeling). The training accuracies on all models approached 1.0, with examples including the LinearSVC achieving a training accuracy of 0.999+; however, the test performance on these models remains very high and stable: in particular, the LinearSVC attained an accuracy of 90.98% and the standard RBF-SVC model (at the 20K subset level) achieved an accuracy of 90.08% and the Random Forest (at 50% of data represented as TF-IDF) attained 89.78%. The small gap between the training and test performance (less than 2–3%) suggests that the classical features of this curated-balanced dataset, such as TF-IDF representation combined with bigram representation and features such as filtering stopwords and frequency capping, captures the lexicon and syntactics of interest for the majority of the instances of these emotions (e.g., love, anger, and joy). There is also a significant drop in accuracy exhibited by the Decision Tree model, which could be indicative of higher sensitivity of that model to the noise and redundancy in the high-dimensional feature space. Although the kernel methods (SVC) maintain a strong theoretical basis for their usage (Cortes & Vapnik, 1995), the marginal improvement gained with kernel methods as compared to the results seen with the linear classifiers within this dataset were similar to what has been reported by Joachims (1998): that in sufficiently large and structured text spaces, the combination of classical feature engineering techniques with appropriate linear classifiers often produce results that equal or exceed that of their kernelized equivalents.

Emotion Distribution Consistency Across Cohorts

The emotion classification counts for actual NSSE open-ended responses as presented in Tables 1 and 2 from 2021 (N=293) and 2019 (N=142) contain a very important finding. Classical models have high internal agreement among models but exhibit a high degree of cohort-dependent bias, while LLMs produce relatively stable predictions between models and cohorts. An example of the differences in emotion classification patterns from Table 1 (the 2021 cohort) is shown below, which indicates how each of the four ML models over-predicts Love (from 37-134 counts) and under-predicts Fear, Anger and Sadness, although the absolute counts of each emotion are not necessarily consistent. The SVC model assigned Love to 61 responses (20.8%) and Anger to 18 responses (6.1%), while in contrast, the Decision Tree model assigned Love to 127 responses (43.3%) and Sadness to only 17 responses (5.8%). The opposite occurred in the 2019 cohort as shown in Table 2; the SVC assigned Joy to 141/142 responses (99.3%) and almost ignored all other emotions. This finding is a strong indicator of label leakage and/or feature dominance, due to an overreliance on certain lexemes, such as "great," "good" etc., without a pragmatic disambiguation process in place.

On the other hand, LLMs show high levels of agreement between model and cohort; they predicted the number of counts of Anger, Sadness and Fear to have the same prediction for all four ML models as less than or equal to 3 counts. Specifically, in the 2021 cohort there was perfect agreement on the emotion Anger as predicted by all four models, with a count of 46 (e.g. GPT-4-Turbo, Claude 3.5, LLaMA-3, and Sonar Pro). In addition, there was also a very small range of predicted counts for the emotion Fear in the 2019 cohort (e.g. 38-39). Overall, these findings support a greater reliability of LLMs than ML models to internalise pragmatic cues (e.g. hedging, intensifiers and discourse markers) and map them back to latent affective states.

Since NSSE responses lack human-annotated emotion labels, absolute accuracy cannot be computed for real-world data. Instead, we treat inter-model agreement as a proxy for reliability—especially valuable in applied settings where ground truth is subjective or unavailable.

**Table 1**: Emotion Prediction Counts on NSSE 2021 Data (N=293)

| **Type** | **Model** | **Joy** | **Surprise** | **Love** | **Anger** | **Sadness** | **Fear** |
|---|---|---|---|---|---|---|---|
| ML | SVC | 134 | 38 | 61 | 18 | 14 | 28 |
| | LinearSVC | 42 | 38 | 108 | 40 | 39 | 26 |
| | RandomForestClassifier | 37 | 48 | 114 | 25 | 46 | 23 |
| | DecisionTreeClassifier | 47 | 31 | 127 | 54 | 17 | 17 |
| | | | | | | | |
| LLMs | OpenAI GPT-4-Turbo | 55 | 41 | 73 | 46 | 50 | 28 |
| | Claude 3.5 | 55 | 40 | 74 | 46 | 49 | 29 |
| | Meta LLaMA 3 (70B) | 58 | 40 | 74 | 46 | 46 | 29 |
| | Perplexity Sonar Pro | 58 | 38 | 72 | 46 | 49 | 30 |

Although the analysis of LLM-based models has demonstrated promising results in terms of scalability and speed. However, it is essential to acknowledge several key limitations that impact their effectiveness. These limitations include challenges related to bias, interpretability, and methodological robustness. For instance, the frequent updates to hosted LLMs pose a challenge for reproducibility over extended periods unless models are version-locked. Additionally, the computational costs associated with running larger LLMs can be significant, especially when compared to traditional machine learning pipelines. Furthermore, LLMs are prone to phenomena such as hallucination, where they may infer meanings or sentiments that are not explicitly present, which introduces potential bias. The lack of a clear ground truth for subjective data, such as sentiment in open-ended responses, further complicates the establishment of definitive accuracy. Moreover, LLM performance is highly dependent on the design of prompts, requiring iterative refinement to achieve optimal results.

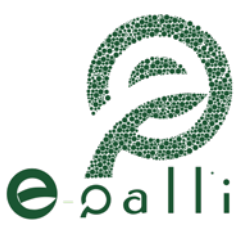

Despite these challenges, there are several avenues for future work. Integrating human validation into the feedback loop can help calibrate and enhance LLM outputs, improving reliability. Additionally, fine-tuning LLMs for specific domains could improve their relevance and accuracy in organizational or thematic contexts. Another promising direction is the development of multi-LLM ensembles, where outputs from multiple models are aggregated to improve decision-making.

**Table 2**: Emotion Prediction Counts on NSSE 2019 Data (N=142)

| Type | Model | Joy | Surprise | Love | Anger | Sadness | Fear |
|---|---|---|---|---|---|---|---|
| ML | SVC | 141 | 36 | 52 | 20 | 7 | 6 |
| | LinearSVC | 57 | 19 | 66 | 55 | 23 | 26 |
| | RandomForestClassifier | 53 | 18 | 92 | 26 | 33 | 20 |
| | DecisionTreeClassifier | 47 | 18 | 101 | 36 | 17 | 27 |
| | | | | | | | |
| LLMs | OpenAI GPT-4-Turbo | 49 | 40 | 52 | 47 | 36 | 38 |
| | Claude 3.5 | 48 | 40 | 51 | 46 | 38 | 39 |
| | Meta LLaMA 3 (70B) | 47 | 40 | 50 | 47 | 40 | 38 |
| | Perplexity Sonar Pro | 47 | 40 | 48 | 48 | 40 | 39 |

**These results raise two key implications:**

1. Classical models are brittle and cohort-sensitive, limiting their utility for longitudinal or cross- institutional survey analysis. Their sensitivity to surface features—notably, lexical valence without context—leads to implausible distributions (e.g., near-universal Joy in 2019), undermining trust in automated coding.
2. LLMs offer convergent validity: high inter-model agreement across architectures (decoder-only, MoE, retrieval-augmented) suggests that emotion attributions reflect shared latent representations rather than idiosyncratic artifacts. While absolute ground-truth labels remain elusive (due to annotator subjectivity;
Bamman & Schneider, 2021), consistency serves as a proxy for reliability—particularly vital when outputs inform high-stakes decisions (e.g., student support resource allocation).

That said, LLMs are not infallible: all four models assign Love more frequently than Joy in NSSE data—a potential artifact of training on social media corpora where "love" is used hyperbolically (e.g., "I love this class!" as enthusiasm, not affective love). Future work should incorporate human-in-the-loop validation and demographic stratification to audit for such biases.

## CONCLUSIONS

This study demonstrates that Large Language Models (LLMS) perform comparatively better than traditional machine learning approaches when analyzing open ended survey responses for sentiments. The traditional machine learning classifiers, such as SVM, Random Forest, and Logistic Regression, require substantial preprocessing, feature engineering, and manually labeled training data; modern LLMs can perform sentiment analysis with little setup required. While large language models (LLMs) excel in cross-cohort consistency, they also have a superior ability to interpret non-literal speech, (i.e., pragmatic cues such as sarcasm). Conversely, traditional models provide the greatest interpretability, represented by feature weights and decision paths, as well as consistent reproducibility when given the same inputs. In high-stakes situations, such as when student sentiment is used to inform policy, it is suggested that when coding with LLMs, a combination of LLM-generated codes and human/ML auditing will provide the highest level of confidence in findings. LLMs face key limitations, bias, hallucination, lack of ground truth for subjective data, and poor reproducibility due to frequent model updates.

Overall, the findings seem to demonstrate that LLMs can help reduce human effort in coding open-text data, increase insight extraction by leaps and bounds, and derive from large-scale qualitative `big data' a minutely nuanced understanding, which marks a major methodological advance for survey analytics.

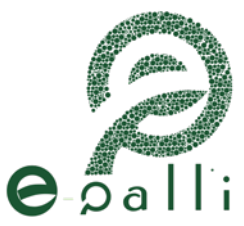

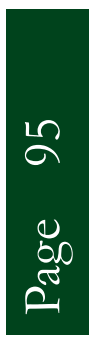

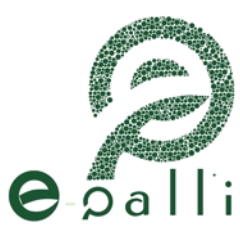